\documentclass[aps,pre,nofootinbib,twocolumn,numbers,showpacs,showkeys]{revtex4-1}
\usepackage[colorlinks=true,citecolor=blue,linkcolor=blue]{hyperref}
\usepackage{natbib}
\usepackage{amssymb}
\usepackage{amsmath}
\usepackage{graphicx}
\usepackage[version=3]{mhchem}

\newcommand{\parder}[3]{\left(\displaystyle\frac{\partial #1}{\partial #2} \right)_{#3}}
\newcommand{\imax}{i_\text{max}}
\newcommand{\nn}{{\tilde n}}
\newcommand{\ww}{{\tilde w}}
\newcommand{\WW}{{\tilde W}}
\newcommand{\dd}{\ensuremath{\mathrm{d}}}

\newcommand{\sol}{\text{sol}}
\newcommand{\gel}{\text{gel}}
\newcommand{\xequ}{i^*}

\begin{document}
\title{Statistical Thermodynamics of Clustered Populations}%
\author{Themis Matsoukas}
\email{matsoukas@engr.psu.edu}
\affiliation{%
Department of Chemical Engineering, Pennsylvania State University, University Park, PA 16802}%

\date{\today}
\begin{abstract}
We present a thermodynamic theory for a generic  population of $M$ individuals distributed into $N$ groups (clusters). We construct the ensemble of all distributions with fixed $M$ and $N$, introduce a selection functional that embodies the physics that governs the population, and obtain the distribution that emerges in the scaling limit as the most probable among all distributions consistent with the given physics. We develop the thermodynamics of the ensemble and  establish a rigorous mapping to regular thermodynamics. We treat the emergence of a so-called ``giant component'' as a formal phase transition and show that the criteria for its emergence are entirely analogous to the equilibrium conditions in molecular systems. We demonstrate the theory by an analytic model and confirm the predictions by Monte Carlo simulation.  

\vspace{20pt}
\end{abstract}
\maketitle

\section{Introduction}
Gibbs's ensemble method \citep{Gibbs:reprint} represents a remarkable success of mathematical physics: it provides the complete macroscopic description of a complex stochastic system and prescribes its conditions of equilibrium and stability in terms of a small number of variables.  The phenomenology of thermodynamics has proven useful in other areas. Systems of evolving discrete populations present a close analogy to thermodynamic microstates and have lead to mathematical treatments that borrow from the language of equilibrium thermodynamics \citep{Vigil:JCIS09,
Berestycki:07,Hendriks:ZPCM85A,Hendriks:ZPCM85B,Spouge:M83_121}. 
Polymer gelation \citep{Flory:JACS41b,Aldous:B99,Lushnikov:PRL04}, shattering in fragmentation \citep{McGrady:PRL87}, the spread of epidemics 
\citep{Pastor-Satorras:PRL01,Newman:PRE02,KrapivskyRednerBenNaim} and the emergence of connectivity in artificial and neural networks \cite{Erdos:MIHA60,Newman:PRL09,Gao:P12,Breskin:PRL06} are all systems that involve the emergence of a giant coherent structure, a process that is commonly discussed in the phenomenology of phase transitions. These diverse physical problems point to a common, yet unclear, link to thermodynamics.  
Jaynes drew a powerful connection between statistical mechanics and information theory \cite{Jaynes:PR57} that suggests a broader applicability of thermodynamics. Jaynes considered the problem of a random variable whose  probability distribution cannot be accessed except through limited data.  He suggested that the unknown distribution can be identified as the one that is most probable among all distributions that are consistent with the available data. This amounts to a constrained maximization of the entropy of the distribution. Given appropriate constraints, practically any distribution may be interpreted as a maximum entropy distribution. The success of the maximum entropy formalism suggests that thermodynamic concepts could be extended outside the realm of interacting particles to any physical problem that involves a distributed population. The unknown distribution would then be determined by constrained maximization of entropy. There is a problem however: there is no obvious connection between the constraints that are required to produce a maximum entropy distribution and the physics from which that distribution emerges. 
For example, depending on the rate law, particle growth by surface deposition may produce a distribution that is Gaussian, if the rate is independent of size, or exponential, if the rate is proportional to particle mass \citep{Matsoukas:PRE06}. Both distributions can be derived by maximum entropy arguments, constrained either by fixing the mean (exponential), or the mean and the variance (Gaussian); there is no physical link, however, between these particular constraints and the rate rate laws from which they arise. 
 
Here we develop a thermodynamic theory that circumvents these difficulties and connects a generic population to the  laws that govern its evolution.  Our approach is motivated by Jaynes in that we view the distribution in the scaling limit as the one that is most probable among all that satisfy the physics of the problem.  We depart from Jaynes in that the distributions of the ensemble are not equally probable but biased by a \textit{selection functional}, a quantity that embodies the physics under which the population evolves. The paper is organized as follows. 
In section \ref{s:ens} we define the microcanonical cluster ensemble and introduce the selection bias and its mathematical properties. In section \ref{s:thl} we pass to the thermodynamic limit, obtain the most probable distribution and derive the fundamental relationships among the primary variables of the ensemble. In section \ref{s:thermo} we discuss the analogies between the cluster ensemble and the familiar thermodynamic ensemble.   In section \ref{s:gel} we develop the thermodynamic criterion for the emergence of a giant cluster (``gel'') and demonstrate the theory with an example, which we solve analytically and test by stochastic simulation. In section \ref{s:discussion} we discuss the findings in the broader context of stochastic processes and finally summarize the conclusions in section \ref{s:conclusions}. 
 
\section{Microcanonical Cluster Ensemble}\label{s:ens}
We begin with a population of $M$ indistinguishable individuals divided into $N$ distinguishable clusters ($N<M$). We define a configuration of masses to be an ordered list of $N$ clusters with total mass $M$ and notate it in vector form as $\mathbf{m} = (m_1,m_2,\cdots,m_N)$.  A configuration is characterized by its distribution $\mathbf{n} =(n_1,n_2,\cdots)$, such that $n_i$ is the number of clusters with $i$ members. We refer to $i$ as the size or mass of the cluster.  The microcanonical ensemble consists of all possible configurations of $M$ members are divided into $N$ clusters.  All distributions of the ensemble satisfy the constraints
\begin{equation}\label{constraints}
   \sum_i n_i = N,\quad
   \sum_i i n_i = M . 
\end{equation}
The maximum possible cluster size in a distribution of the $(M,N)$ ensemble is $M-N+1$, but if we adopt the convention $n_i=0$ for all $i>M-N+1$, the limits in the summations may be assumed to run from $i=$1 to $\infty$ and will not be explicitly shown.
Each distribution is associated with a multiplicity factor that represents the number of configurations that have the same distribution. This is equal to the number of permutations in the order of cluster masses in the configuration and is given by the multinomial coefficient,
\begin{equation}
   \mathbf{n!} = \frac{N}{n_1!\,n_2\!\cdots}. 
\end{equation}
The log of the multiplicity factor is the entropy of distribution:
\begin{equation}
   S = \log\mathbf{n!} . 
\end{equation}
In the Stirling approximation this reverts to the familiar functional.  We now introduce the \textit{selection bias} $W(\mathbf{n})$, a functional of $\mathbf{n}$ that biases selection such that the probability of distribution $\mathbf{n}$ is proportional not only to its multiplicity $\mathbf{n!}$ but also to its selection bias $W(\mathbf{n})$:  
\begin{equation}\label{P_n}
   P(\mathbf{n}) = \mathbf{n!}\frac{W(\mathbf{n})}{\Omega_{M,N}} .
\end{equation}
Here $\Omega_{MN}$ is the partition function and satisfies the normalization condition,
\begin{equation}\label{Omega:norm}
   \Omega_{M,N} = \sum_\mathbf{n}\mathbf{n!}\, W(\mathbf{n}) , 
\end{equation}
with the summation taken over all distributions of the ensemble. 
The selection bias is a fundamental property of the cluster ensemble and embodies the physics of the problem. Here, the bias will remain general and unspecified; the only  condition we impose is that it must produce distributions with  proper extensive behavior in the thermodynamic limit. This requires $\log W$ to be homogeneous in $\mathbf{n}$ with degree 1 (see the Appendix for details). It then follows from Euler's theorem that $\log W$ is a linear combination of its derivatives with respect to $n_i$: 
\begin{equation}\label{logW}
   \log W(\mathbf{n}) = 
   \sum_i n_i \parder{\log W(\mathbf{n})}{n_i}{n_j} \equiv
   \sum_i n_i\log w_i .
\end{equation}
The derivatives $\log w_i$ are an important element of the theory; they represent the contribution of cluster size $i$ to $\log W$ and will be referred to as \textit{cluster bias}.  If the log of the selection bias is a \textit{linear} functional of $\mathbf{n}$ \citep{GelfandFromin}, then the $w_i$ are intrinsic functions of cluster size $i$, and thus independent of the distribution itself. A special case of linear bias is $W(\mathbf{n})=1$ ($\log w_i=0$), which gives equal weight to all distributions (unbiased ensemble). In the general case the $w_i$'s will depend not only on $i$ but also on the distribution $\mathbf{n}$ to which they refer, as Eq.\ (\ref{logW}) implies. 

The microcanonical ensemble defined here is closely related to that discussed in \citet{Pitman:06} and in \citet{Berestycki:07} in the context of integer partitions. The main difference is in the natural multiplicity of distributions in the ensemble, which depends on whether cluster configurations are taken to be ordered (as in the cluster ensemble), or not (as in Refs.\ \citep{Pitman:06,Berestycki:07}). Such differences can be reconciled between the two ensembles.  References \citep{Pitman:06,Berestycki:07} also present solutions for special cases of linear bias functionals that arise in coagulation and fragmentation. Here our interest is not in specific solutions but rather in the fundamental relationships between variables of the cluster ensemble. 
 
\subsection*{Exchange Reactions}
Suppose that two clusters in distribution $\mathbf{n}$ with sizes $i$ and $j$, respectively, exchange members to produce two new clusters with sizes $k$ and $l$ such that $i+j=k+l$.  This binary exchange process converts one distribution of the ensemble into another and can be represented by the reversible reaction,
\begin{equation}
    \ce{\mathbf{n} <=>[K]\mathbf{n'}}
\end{equation}
If the equilibrium constant is set to
\begin{equation}\label{prob_transition}
   K_{\mathbf{n}\rightleftharpoons\mathbf{n'}} =\frac{W(\mathbf{\mathbf{n'}})}{W(\mathbf{n})} ,  
\end{equation}
the process will then produce an ensemble of distributions whose probability is given by Eq.\ (\ref{P_n}). The exchange reaction, allows us to associate the cluster ensemble, a static collection of distributions, with a dynamic ensemble that obeys detailed balance with respect to exchange reactions. The practical implication is that we may sample the ensemble by Metropolis Monte Carlo simulation of binary exchange reactions with acceptance probability proportional to $W(\mathbf{n'})/W (\mathbf{n})$.\footnote{This extends to any type of exchange reaction, not necessarily binary with respect to clusters, as long as both $M$ and $N$ are conserved. Binary reactions are convenient because they are the simplest to implement by simulation. }

\section{Thermodynamic Limit}\label{s:thl}
We now pass to the thermodynamic limit on the premise that when $M$ and $N$ are large the ensemble reduces to a single distribution, $\nn$. From Eq.\ (\ref{Omega:norm}) we then have
\begin{equation}\label{Omega_S_W}
   \log\Omega \to \tilde S + \log \WW , 
\end{equation}
where $\tilde S = S(\mathbf{\nn})$ is the entropy of the most probable distribution and $\log\WW=\log W(\mathbf{\nn})$ is its log-bias. 
The most probable distribution is obtained by standard Lagrange maximization, and the result is
\begin{equation}\label{mpd}
   \frac{\nn_i}{N} = \ww_i\frac{e^{-\beta i}}{q} ,
\end{equation}
where  $\log \ww_i$ is the cluster bias in the most probable distribution.  The Lagrange multipliers $\beta$ and $\log q$ correspond to the two constraints in Eq.\ (\ref{constraints}). They are given in terms of the ratio $M/N$ in implicit form by 
\begin{align}
\label{beta:wi}
   \frac{M}{N}&= \sum i\,\ww_i e^{-\beta i}\Big/\sum \ww_i e^{-\beta i},\\
\label{q:wi}
   q          &= \sum \ww_i e^{-\beta i},
\end{align}
which we obtain by inserting the most probable distribution into the constraints. We now take the log of the most probable distribution, multiply by $\nn_i$, and perform the summation (see Appendix): 
\begin{equation}\label{PF_fundamental} 
   \log\Omega_{M,N} = \beta M + (\log q)\, N ,
\end{equation}
This result of remarkable simplicity represents a fundamental relationship between the primary variables of the ensemble.  Since $S$ and $\log W$ are extensive in $M$ and $N$, it follows from Eq.\ (\ref{Omega_S_W}) that so is $\log\Omega$. In other words, $\log\Omega$ is homogeneous in $M$ and $N$, with degree 1. By Euler's theorem we have
\begin{equation}\label{beta_q_fundamental}
   \beta = \parder{\log\Omega}{M}{N};\quad
   \log q = \parder{\log\Omega}{N}{M} , 
\end{equation}
and  
\begin{equation}\label{dPF_fundamental}
   \dd\log\Omega_{MN} = \beta\, \dd M + (\log q)\,\dd N . 
\end{equation}
Equation (\ref{beta_q_fundamental}) identifies $\beta$ and $q$ as the partial derivatives of the microcanonical log-partition function with respect to its extensive variables, while Eq.\ (\ref{dPF_fundamental}) governs the  evolution of a population under a quasistatic change of state $(\dd M,\dd N)$.

\subsection*{Canonical Ensemble}
To complete the theory we derive the statistics of the canonical ensemble.  We start with a large microcanonical ensemble of $M'$ individuals distributed into $N'$ clusters. The ensemble is characterized by fixed $\beta$, $q$. We sample randomly  $N$ clusters ($N\ll N'$) from a configuration of this microcanonical pool and seek the probability of distribution $\mathbf{n}$ that is sampled in this manner.  In the thermodynamic limit the probability to pick a cluster of size $i$ is given by the relative frequency of cluster size $i$  ensemble, $P_i=w_i \exp(-\beta i)/q$. For small $N$ relative to $N'$ the probability to sample distribution $\mathbf{n}=(n_1, n_2,\cdots)$ is $\mathbf{n!} P_1^{n_1} P_2^{n_2}\cdots$. Upon expanding the product the result becomes 
\begin{equation}\label{P:canonical}
   P(\mathbf{n}) = \mathbf{n!}W(\mathbf{n}) \frac{e^{-\beta M}}{q^N},
\end{equation}
where $M$ is the total mass (number of members) in the sampled distribution. Strictly, this derivation applies to linear bias because it assumes the $w_i$ in Eq.\ (\ref{P:canonical}) to be the same for all distributions. Nonetheless, a sharply peaked ensemble in the vicinity of the most probable distribution is adequately described by a linearized bias and the result applies to general bias in the scaling limit. 

The canonical partition function is the sum of the canonical weights $\mathbf{n!} W(\mathbf{n!})\exp(-\beta M)$. Since  probabilities in Eq.\ (\ref{P:canonical}) are properly normalized, by summation over all $\mathbf{n}$ on both sides we obtain  
\begin{equation}\label{logQ}
   Q = q^{N}. 
\end{equation}
Here we recognize $\log q$ as the intensive form of the logarithm of the canonical partition function, $(\log Q)/N$.

\section{Connection to Thermodynamics}\label{s:thermo}
An analogy to molecular thermodynamics now emerges. Compare Eq.\ (\ref{dPF_fundamental}) with the familiar thermodynamic relationship in the $nVE$ ensemble,
\begin{equation*}
   \frac{\dd S}{k}  = \frac{\dd E}{kT} + \frac{p\,\dd V}{kT}  
                     - \frac{\mu\,\dd n}{kT}  .  
\end{equation*}
In both cases, on the left-hand side we have the differential of a quantity  whose maximization defines the equilibrium state, expressed on the right-hand side in terms of a set of extensive variables that fix the macroscopic state of the system. A one-to-one correspondence can be drawn between the two ensembles, as shown in Table \ref{tbl:thermo} (to complete the analogy we must also set $\mu=0$). 
%
\begin{table}
\caption{Correspondence between properties of the cluster ensemble and thermodynamics. Other mappings are possible. }
\begin{equation*}
\begin{array}{ccc}
\text{cluster ensemble} & & \text{Thermodynamics} \\\hline
\log\Omega &\to& (\text{thermodynamic Entropy})/k\\
M &\to & \text{Energy} \\
\beta &\to & 1/kT \\
N &\to & \text{Volume} \\
\log q &\to & p/kT \\
\hline
\end{array} 
\end{equation*}
\label{tbl:thermo}
\end{table}
%
Notice that the maximized quantity of the cluster ensemble is not the entropy of distribution, as in statistical mechanics, but the microcanonical partition function.  In statistical mechanics the selection bias is set by the postulate of equal a priori probabilities. This corresponds to uniform bias $W=1$ in our theory. Equation (\ref{Omega_S_W}) then gives $\log\Omega=S$ and thus we recover the familiar $nVE$ ensemble.  In general, $\log\Omega$ and entropy in the cluster ensemble are distinct properties, and between the two, it is the partition function that is of fundamental importance. 

The thermodynamic mapping shown here is not unique as one may choose to associate $\beta$ with $-\mu/kT$, for example; this mapping is implied in \citet{Ziff:JPMG83}, who refer to the quantity $e^{-\beta}$ as ``fugacity'' (symbol $\xi$ in the original reference). In this sense, the correspondence between variables of the cluster ensemble and thermodynamics should not be taken literally but ought be viewed as a mathematical mapping that allows one to obtain relationships between variables of the cluster ensemble by translation of the corresponding result in thermodynamics. For example, by direct analogy to the thermodynamic result, $C_V>0$, we may write, $(\partial M/\partial)_N\beta<0$, a result that we can independently obtain by stability analysis on $\log\Omega$.  
Still,  the mapping in Table \ref{tbl:thermo} has a certain intuitive appeal. For example, a process may be viewed as ``compression'' if it causes the average size to increase (we imagine $N$ to decrease at constant $M$), or ``expansion,'' if the average size decreases.

\section{A Phase Transition: the giant cluster}\label{s:gel}
If populations are thermodynamic entities, do they undergo the equivalent of phase equilibrium? In the context of the cluster ensemble, a ``phase'' is a distinct distribution, and ``phase equilibrium'' refers to the coexistence of two separate distributions within the same population, such that they both remain stable under exchange of members. We address this problem by considering the emergence of a giant cluster, namely, the transition from a population of finite clusters to one that contains one cluster that is of the order of $M$.  The emergence of the giant component has long invited an analogy to condensation, but now we are in position to make a rigorous connection. 
 
When $M$ individuals are placed into $N$ clusters, the maximum possible cluster size is $\imax=M-N+1$. The size range $\imax/2<i\leq\imax$ is special: it can accommodate \textit{at most one cluster}; the mass balance is not satisfied otherwise. A cluster in this region contains a \textit{finite} fraction of the total mass (its size is of the the order of $M$) but a vanishingly small fraction of the total number of clusters. We refer to clusters in this range as the ``gel'' phase and to all others as ``sol'' phase. 
Such system distributes members between the two phases so that the partition function of the combined system is maximized under the constraints $M_\sol+M_\gel=M$ and $N_\sol+N_\gel=N$ with  $N_\gel=1$. This variational principle leads to the condition 
\begin{equation}\label{isoT}
   \parder{\log\Omega_\sol}{M_\sol}{N_\sol} = \parder{\log\Omega_\gel}{M_\gel}{N_\gel} \equiv \beta. 
\end{equation}
This result is obtained by maximizing the partition function of the  two-phase system  (see Appendix) and establishes the condition of thermal equilibrium between the phases. Since the number of clusters in each phase is fixed,  $q$ is \textit{not} required to equilibrate. The emergence of a giant cluster is akin to osmotic equilibrium.

\subsection*{A Gelling Selection Bias}
We demonstrate the application of the theory using a selection bias that  can be solved  analytically. We take the cluster bias to be $w_i = i^{-3}$, which corresponds to the selection functional,  
\begin{equation*}
   W(\mathbf{n}) = \prod_i i^{-3n_i} . 
\end{equation*}
Let us examine whether it is possible to construct a two-phase solution for this bias. 
The distribution of the sol phase is 
\begin{equation}\label{i3:mpd}
   \left(\frac{n_i}{N}\right)^\sol = i^{-3}\frac{e^{-\beta i}}{q},
\end{equation}
with $\beta$ and $q$ given by 
\begin{align}
\label{i3:beta}
   \frac{M}{N} &= \frac{\text{Li}_2\left(e^{-\beta}\right)}
                         {\text{Li}_3\left(e^{-\beta }\right)}\\
\label{i3:q}
   q &=\text{Li}_3\left(e^{-\beta }\right) .
\end{align}
where $\text{Li}_n(x)$ is the polylogarithm function. These are obtained From Eqs.\ (\ref{beta:wi}) and (\ref{q:wi}) by letting the upper limit to go to infinity (scaling limit). To obtain $\beta$ and $q$ we solve Eqs.\ (\ref{i3:beta}) and (\ref{i3:q}) numerically for the given $M/N$. The sol distribution is therefore completely defined if $M/N$ is fixed. 

Suppose the system forms a two-phase state that consists of a gel cluster with mass $M_\gel$ in a equilibrium with sol ($M_\sol=M-M_\gel$, $N_\sol=N-1\to N$). For the linear bias functional, the partition function of the gel-phase is $\Omega_\gel=w(M_\gel)$, where $w(M_\gel)$ is the cluster bias of the giant cluster. This follows from Eq.\ (\ref{Omega_S_W}) and the fact that the gel phase consists of a single cluster ($S_\gel=0$). 
%
\begin{table}
\caption{Phase behavior of linear cluster bias $w_i=i^{-3}$. The system forms a giant cluster at $M/N=1.368$. }
\label{tbl:phase}
\begin{equation*}
\renewcommand{\arraystretch}{1}
\begin{array}{c| cc ccc}
\multicolumn{1}{c}{}
   &\makebox[80pt][c]{$M/N<1.368$}  
   &\makebox[80pt][c]{$M/N\geq1.368$} 
\\\cline{2-3}
\multicolumn{1}{c}{}
   &\text{single sol}   
   &\text{sol+gel} \\
\hline
\text{sol distribution}
   & \displaystyle\frac{n_i}{N} =  \frac{e^{-\beta i}}{q}i^{-3}
     \vphantom{\rule[-0mm]{0cm}{0.75cm}}
   & \displaystyle\frac{n_i}{N} = 0.832\, i^{-3}
\\
\text{gel distribution} 
   & - 
   & n_i^\gel = \delta_{i,M_\gel}
\\
\beta
   & \text{Eq.\ \ref{i3:beta}}
   & 0
\\
q
   &\text{Eq.\ \ref{i3:q}}
   & 1.202
\\
\text{gel fraction} 
   & 0 
   & \displaystyle\frac{M_\gel}{M} = 1 - 1.368\frac{M}{N}
     \vphantom{\rule[-0.55cm]{0cm}{0.55cm}}
\\
\hline
\end{array} 
\end{equation*}
\end{table}
%
Its temperature is
\begin{equation}\label{bgiant}
   \beta^\gel = \frac{\dd\log w_i}{\dd i}\Big|_{M_\gel} = -3/M_\gel,
\end{equation}
and in the scaling limit, $\beta^\gel\to 0$. 
The corresponding equilibrium sol phase is obtained from Eq.\ (\ref{i3:mpd}) with $\beta=0$, $q=1.202$. This produces the power-law distribution
\begin{equation}\label{i3:mpd:equ}
   \left(\frac{\nn_i}{N}\right)^* = 0.832 i^{-3},
\end{equation}
whose mean cluster size is $i^*=1.368$.  The phase behavior is now fully determined. For $M/N<i^* = 1.368$ we have a single sol phase whose distribution is,
\begin{equation}\label{mpd:theory}
   \frac{n_i}{N} = \frac{ i^{-3} e^{-i\beta}}{\text{Li}_3\left(e^{-\beta}\right)} . 
\end{equation}
%
\begin{figure}[t]
\begin{center}
\includegraphics[width=\columnwidth]{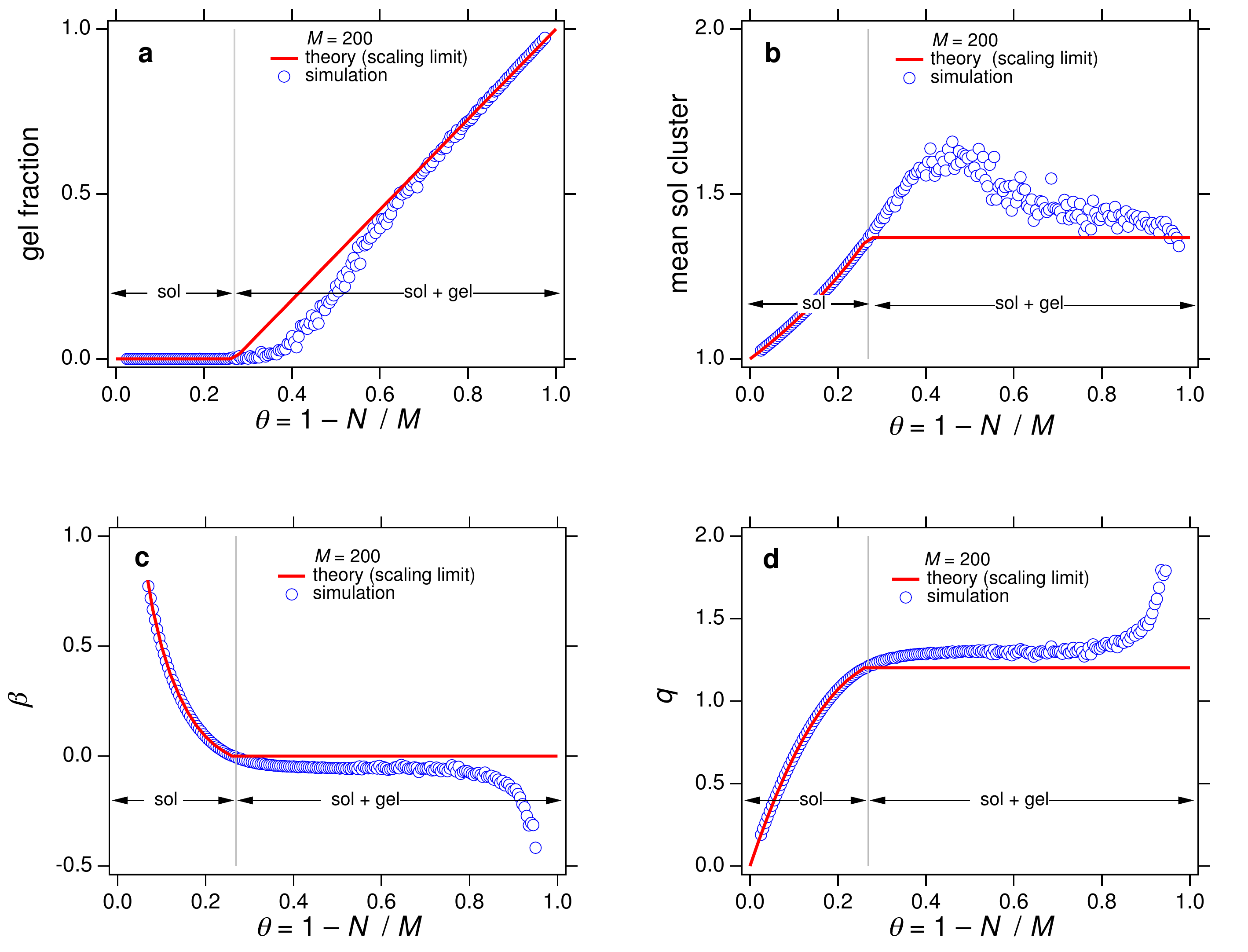}
\end{center}
\caption{(Color online) Phase diagram of a gelling system with cluster bias $w_i=i^{-3}$: (a) gel fraction, (b) mean sol cluster, (c) temperature $\beta$, and (d) pressure $q$, as a function of the progress variable $\theta=1-N/M$. The simulations are conducted with $M=200$. The theoretical lines are based on the equations in Table \ref{tbl:phase}, which assume an infinite system. 
}
\label{fig1}
\end{figure}%
%
If $M/N$ is larger than the equilibrium sol cluster,  the system splits in two phases: a sol phase whose distribution is given in Eq.\ (\ref{i3:mpd:equ}), and one giant cluster whose fraction is given by the mass balance condition (see Appendix)
\begin{equation}\label{phi}
   \phi_\gel = 1-\xequ \left(\frac{N}{M}\right),  
\end{equation}
where $\xequ=1.368$ is the equilibrium mean sol cluster.  Notice that the normalized equilibrium sol distribution $(n_i/N)_\text{sol}$ remains constant in the entire sol/gel region.  The conversion of the sol phase into the gel along the path $N\to 1$ at fixed $M$ occurs isothermally until the sol disappears completely. 

We test these predictions by Monte Carlo simulation using the method of exchange reactions on a finite population with $M=200$ members. We begin with a configuration (ordered list) of $N$ cluster masses with total mass $M$;  the initial distribution is approximately uniform but any other distribution may be used. At each step we pick two clusters at random, merge them into a single cluster, then break it up randomly in two. The new configuration is accepted with probability proportional to the equilibrium constant in Eq.\ (\ref{prob_transition}) and the process is repeated until the system reaches equilibrium.  The mean distribution is calculated as an average of $1$ to $4\times10^6$ realizations. The simulation is repeated with $N$ ranging from $M-1$ to 2.

\begin{figure}
\begin{center}
\includegraphics[width=2.5in]{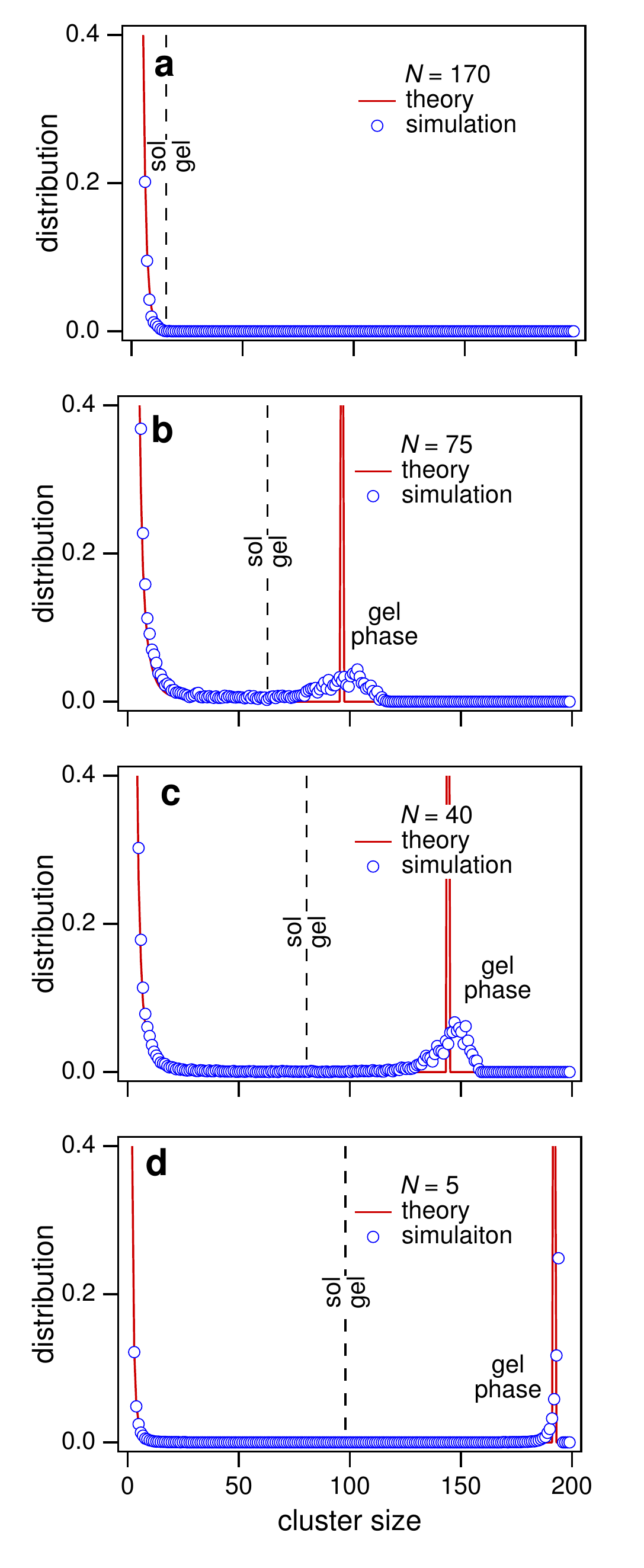}%
\end{center}
\caption{(Color online) Selected distributions for cluster bias $w_i=i^{-3}$ in a finite population with $M=200$ members. The giant cluster appears near $N^*=146$. The shaded area is the simulated distribution and the solid line is theory.  The dashed line marks the phase boundary between the sol phase and the gel phase. }
\label{fig2}
\end{figure}
First we analyze the sol and gel phases obtained in the simulation. We calculate the gel fraction, $\phi_\gel$, as the fraction of the total mass that resides in the region $i>(M-N+1)/2$. The mean cluster size in the sol ($\bar i$) is calculated as the average mass in the region $1\leq 1 < (M-N+1)/2$, and the parameters $\beta$ and $q$ are obtained from the finite version of Eqs.\ (\ref{i3:beta}) and (\ref{i3:q}). These results ($\phi$, $\bar i$, $\beta$ and $q$) are shown in Fig.\ \ref{fig1} as a function of the progress variable $\theta = 1-N/M$, where $\theta=0$ refers to a fully dispersed population (perfect sol), and $\theta=1$ a fully gelled population (all particles have joined the giant gel).  
The simulations are in general agreement with the theory though discrepancies are noted due to finite-size effects. The gel point is predicted at $\theta^*=1-1/i^*=0.2690$ ($N^*=146$), while the simulation gives $\theta^*\approx 0.3$ ($N^*\approx 140$). The gel fraction is seen to be a linear function of $\theta$ with small deviations seen near the gel point. The mean cluster in the post-gel region converges to the predicted   value $i^*=1.368$; its small value makes it quite more sensitive to finite-size effects. Recall that in these simulations $N$ is varied at constant $M$. In thermodynamics terms, in the direction of increasing $\theta$ the system undergoes compression. Accordingly its temperature $1/\beta$ increases, and so does its pressure $\log q$.

Sample distributions are shown in Fig.\ \ref{fig2}. At $N=170>N^*$ the population consists of a single sol distribution that decays monotonically within the sol region and is in excellent agreement with Eq.\ (\ref{i3:mpd}). At $N=75<N^*$ the population is compressed above its gel point, a gel phase is present, and the sol distribution is now given by Eq.\ (\ref{i3:mpd:equ}). As $N$ is reduced further, the sol phase ``shrinks in place,'' i.e., the number of sol clusters decreases while the normalized distribution remains fixed, and the giant cluster moves to larger size as more particles are converted from the sol. The theoretical distribution of the gel phase is a Kronecker delta at $M_\gel=M(1-1.368\,N/M)$. The apparent distribution of gel clusters represents not the relative number of gel clusters in a typical distribution (there is at most one cluster in this range) but fluctuations in the position of the giant cluster from one configuration to the next.

\begin{figure}
\begin{center}
\includegraphics[width=2.25in]{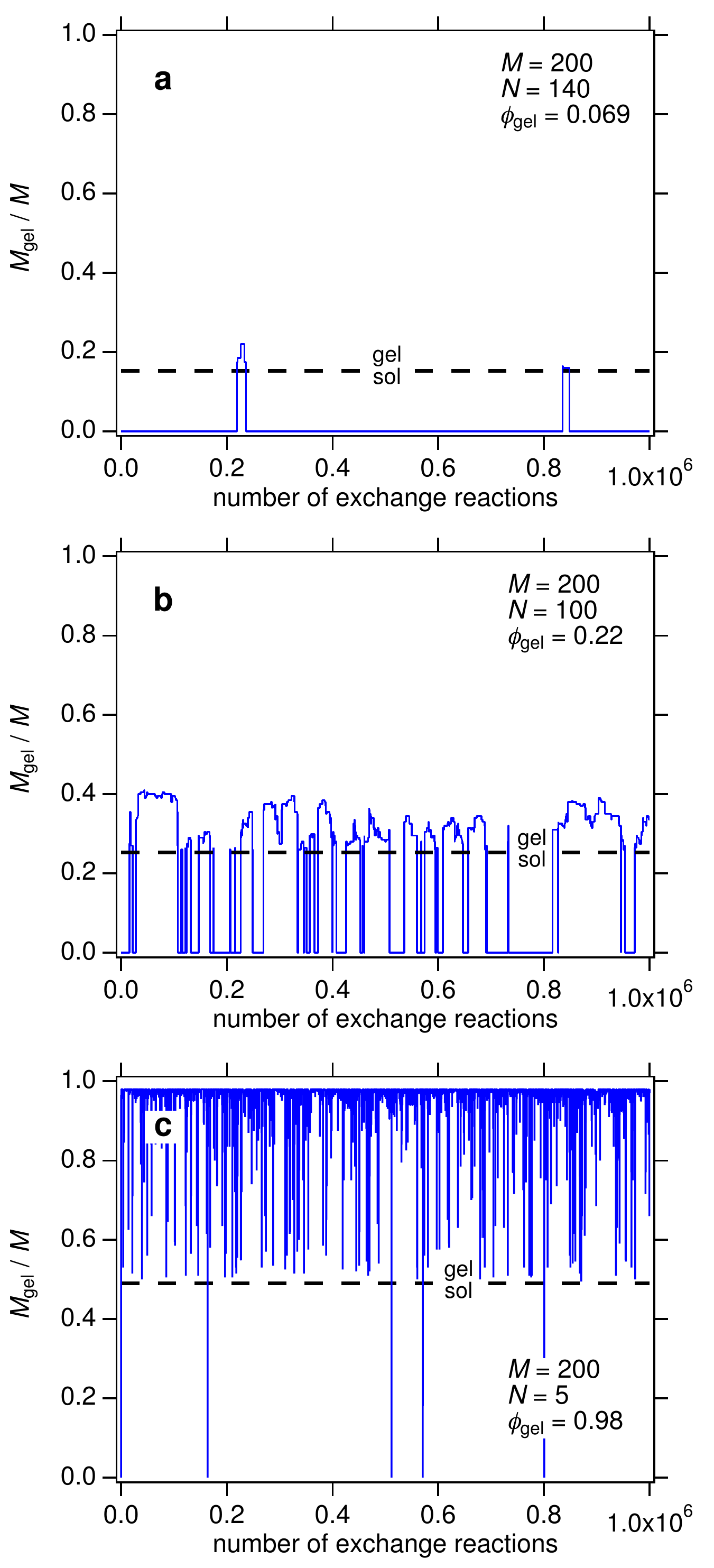}
\end{center}
\caption{(Color online) Fluctuations of the gel cluster with cluster bias $w_i=i^{-3}$ over the course of exchange reactions in a population with $M=200$ members. (a) $N=140$ (right about the gel point); (b) $N=100$ (above the gel point); (c) $N=5$ (very close to complete gelation). The dashed line marks the boundary between the sol and the gel phase. When the gel cluster falls below this line, it enters the sol phase and the gel fraction drops to zero.  }
\label{fig3}
\end{figure}

We examine these fluctuations in more detail in Fig.\ \ref{fig3}, which shows the mass fraction of the giant cluster, $M_\gel/M$, within individual configurations of the ensemble and how it evolves as a result of exchange reactions. The dashed line marks the sol-gel boundary. A cluster above this line is a gel cluster; if the  maximum cluster falls below the dashed line (sol region), no gel cluster exists, i.e. the giant cluster evaporates into the sol and its mass is zero. Crossings above and below the dashed line indicate condensation/evaporation, respectively, of the gel cluster.  These fluctuations therefore reflect transfer between the two phases. 
At $N=140$ ($\theta=0.3$) the system is close to the  gel point. The nucleation of the gel phase is indicated by short-lived excursions into the gel region. The gel fraction, calculated over all states (whether they contain a gel cluster or not) is $\phi_\gel=0.069$. Even though the giant cluster appears abruptly in the size range $i>\imax=30$, the ensemble-average gel fraction at the gel point is continuous. 
At $N=100$ ($\theta=0.5$) the system is above the gel point. Most states contain a gel cluster that makes frequent crossings into the sol region. The mean gel fraction is $\phi_\gel=0.22$. 
At $N=4$ the system is very close to complete gelation with most of its mass residing on the giant cluster ($\phi_\gel = 0.98$). Even at this highly gelled state, fluctuations reach the sol-gel boundary and occasionally the giant cluster evaporates but these excursions are both less frequent and short-lived. 

\begin{figure}
\begin{center}
\includegraphics[width=2.5in]{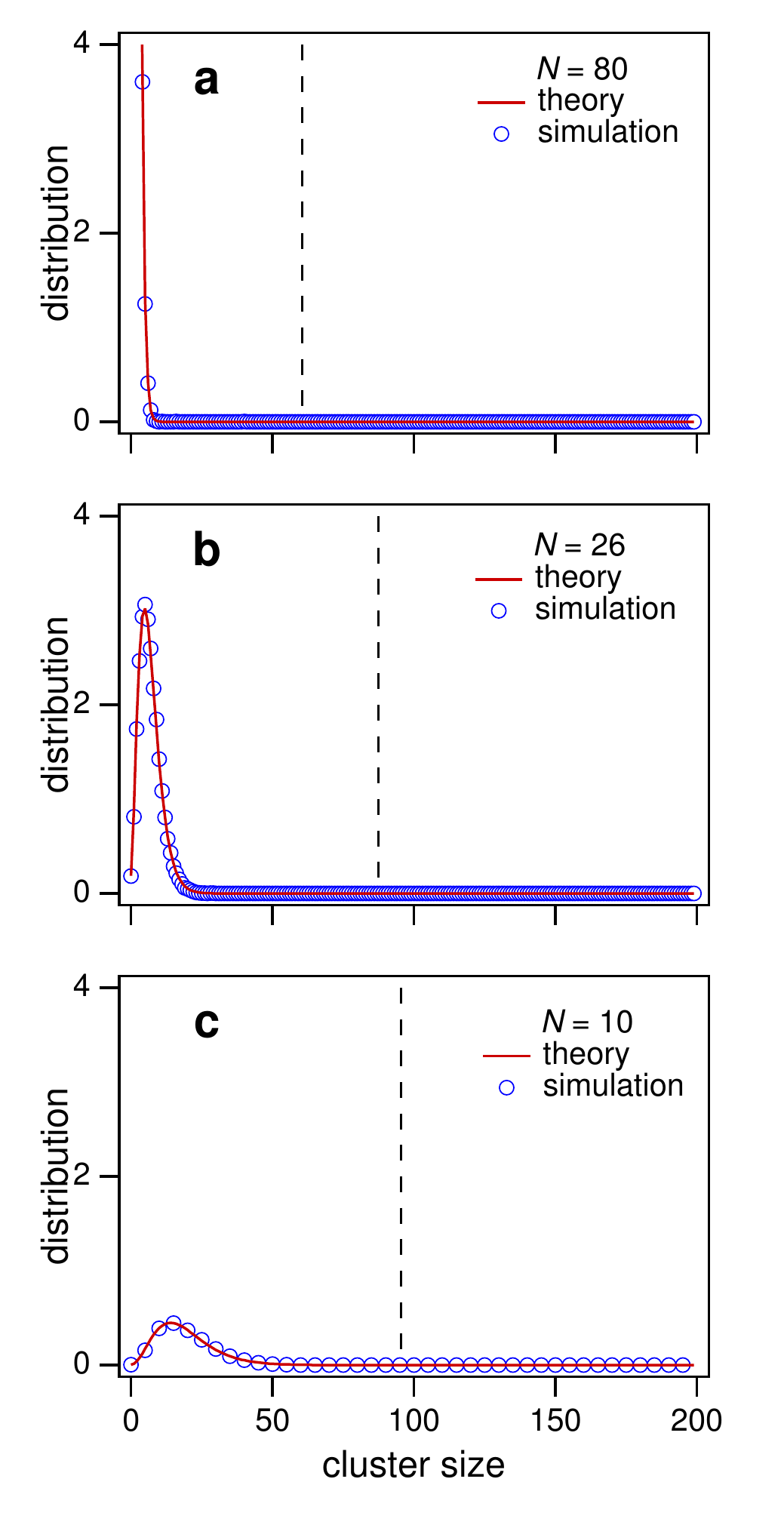}
\end{center}
\caption{(Color online) The cluster bias $w_i = i^{3}$ results in a stable single-phase population for all $N$ (no gel phase).  The shaded area is the simulated distribution and the solid line is theory. The dashed line marks the phase boundary between sol and the region of the giant cluster.}
\label{fig4}
\end{figure}
The phase behavior of the gelling bias studied here bears remarkable similarity to Stockmayer's theory of gelation \citep{Stockmayer:JCP43}. Stockmayer constructs an ensemble of polymer chains, similar to our cluster ensemble, and selects chains in proportion to the number of ways that $i$ monomers can be combined to form an $i$-mer. He obtains the most probable distribution by maximizing the partition function and finds the gel fraction to be a linear function of the fraction of unreacted bonds, a quantity analogous to our $\theta$. While Stockmayer raises numerous  analogies between his gelation model and the thermodynamics of phase transitions, he obtains phase diagram by analysis of the convergence radius of the cluster distribution, not by thermodynamics, and identifies the gel point as the limiting condition that guarantees convergence of the moments of order 0 and 1. We can now make the thermodynamic connection rigorous.
The cluster bias in Stockmayer's model is\footnote{Notice that the $w_i$ in Stockmayer's notation refers to $i!w_i$ in ours.} \citep{Stockmayer:JCP43}
\begin{equation}\label{stockmayer}
    w_i =  \frac{f!(f i - i)!}{i! \,(f i - 2 i+2)!} ,
\end{equation}
where $f$ is the functionality of the monomer. For $f=3$ (the general case $f>3$ can be treated similarly) this is functionally equivalent to the cluster bias (see the Appendix)
\begin{equation}
   w_i = i^{-5/2} . 
\end{equation}
This is of the same power-law type as the gelling bias discussed above and leads to the same isothermal condition at the sol-gel point,  $\beta^\gel=\beta^\sol=0$. This condition also defines the radius of convergence of the summations in Eqs.\ (\ref{beta:wi}) and (\ref{q:wi}): the first and second moments of $\nn_i$ converge in the region $\beta\geq 0$, while the second moment converges in $\beta>0$ but diverges at the gel point. This is precisely the method by which Stockmayer obtained the gel point. Here we arrived at the same result by strict thermodynamic analysis. 

Not every selection bias leads to a gelling population. Consider the cluster bias $w_i=i^3$, the inverse of the case discussed above, whose sol distribution is
\begin{equation*}
  n_i/N = i^{3} e^{-\beta i} / q . 
\end{equation*}
The temperature of the giant cluster, if one forms, is $\beta^\gel = 3/M_\gel$. Again, in the scaling limit we obtain $\beta^\gel\to 0$ but now the corresponding sol distribution diverges and cannot produce a sol that satisfies the two constraints of the ensemble.  It is not possible to construct an equilibrium two-phase mixture this case. Even though this selection bias favors distributions with heavy tails, these are fully contained in the sol region. This prediction is confirmed by Monte Carlo simulation (Fig.\ \ref{fig4}), which demonstrates that the system forms a single sol distribution for all $M/N$.

\section{Discussion}\label{s:discussion}
The main element of the theory is the selection bias $W$, which alone determines the distribution in the scaling limit. The connection to statistical mechanics may now be seen more clearly. In statistical mechanics we begin with the postulate of equal a priori probabilities, which corresponds to the unbiased ensemble with $W=1$, and obtain the exponential distribution of microstates as the most probable distribution in the in the scaling limit. By allowing the selection bias to be an arbitrary functional of the distribution $\mathbf{n}$, it is possible to  obtain \textit{any} distribution in the scaling limit. Conversely, any distribution of $M$ members clustered in $N$ groups can be associated with an equilibrium ensemble that obeys thermodynamics, provided that the corresponding selection bias is identified. The broader implication is that stochastic variables may be treated in the formalism of thermodynamics by associating its probability density function with the corresponding equilibrium ensemble. In this sense, thermodynamics should be viewed as a probabilistic calculus that is applicable not only to systems of physical particles, but to stochastic processes in general. Complex problems such as gelation, percolation and network connectivity may then be treated as formal phase transitions that obey thermodynamics. 
The crux of the problem then is the determination of the selection bias. This must be done on a case-by-case basis, based on the physical laws that govern the system at hand. In the Stockmayer model, for example, the selection bias arises from a combinatorial model for polymer chains. In populations undergoing dynamic transformations, it is determined by the corresponding rate laws. Examples will be discussed elsewhere.

We close with a final note on dynamics. Although time is not an explicit element of the theory, the connection to dynamics is provided by Eq.\ (\ref{dPF_fundamental}). This equation describes the evolution of the ensemble along a quasi-static trajectory, which we define as a path along which the selection bias remains invariant, while $M$ and $N$ change due to dynamic processes in the population. Essentially, Eq.\ (\ref{dPF_fundamental}) gives the evolution of the system in terms of the mean cluster size $M/N$ rather than time.  At fixed $(M,N)$ the system is at ``equilibrium,'' which is understood to mean that the ensemble relaxes to the most probable distribution that corresponds to the present values of $M$ and $N$. Even if the system is on an irreversible path, i.e., unidirectional in time, the state at fixed $(M,N)$ may be described using the calculus of equilibrium thermodynamics. The validity of this approach is confirmed in Fig.\ \ref{fig1}: in moving from $\theta=0$ to $\theta=1$ we follow a polymerization reaction that leads to the irreversible formation of a gel.

\section{Conclusions}\label{s:conclusions}
In summary, we have presented a thermodynamic theory for a generic population of $M$ individuals clustered into $N$ groups. The distribution of this population is viewed as the most probable distribution that emerges among all possible partitions of $M$ into $N$ elements, under a functional that biases the selection of individual partitions.  By associating the population with a corresponding ensemble we are able to express the distribution in terms of thermodynamic quantities (partition function, ``temperature,'' ``pressure'') that obey relationships analogous to those in molecular systems. An important result of the theory is a thermodynamic criterion (maximization of the partition function) to determine whether a giant cluster (gel phase) forms. Although it is common in the literature to refer to the emergence of a giant component as a phase transition, the theory presented here associates this process, for the first time,  with the maximization of a thermodynamic functional. This unlocks the toolbox of thermodynamics and makes it available to the study of population balances and stochastic phenomena in general.

\appendix
\begin{center}
\bfseries APPENDIX
\end{center}
\section{Derivations and Additional Notes}\label{app:derivations}
\subsection{Homogeneity Condition Eq.\ (\ref{logW})}
Here we show that in order for the ensemble to have proper extensive behavior in the thermodynamic limit, the log of the selection bias must be a homogeneous function of $n_i$ with degree 1. Proper extensive behavior means that if $M$ and $N$ are both increased by a factor $\lambda$, the number of clusters $\nn_i$ in the most probable distribution must increase by the same factor. Equivalently, the ratio $\nn_i/N$ must be an intensive property, i.e., a function of $M/N$. The most probable distribution

We start with the relationship between $\beta$, $q$ and $M/N$ in Eqs.\ (\ref{beta:wi}), (\ref{q:wi}), which we rewrite as
\begin{align}
\label{app:homo_b}
   \frac{M}{N} &= \sum i e^{-\beta i + \log\ww_i}
   \Big/
   \sum e^{-\beta i + \log\ww_i}, \\
\label{app:homo_q}
   q           &= \sum e^{-\beta i + \log\ww_i} . 
\end{align}
The ratio $\nn_i/N$ will be intensive only if $\beta$, $q$, and $\log\ww_i$ are intensive; in turn, $\beta$ and $q$ will be intensive only if the $\log\ww_i$ are intensive. This implies that if a distribution $\mathbf{n}$ is multiplied by $\lambda$, which will result in all cluster numbers being multiplied by $\lambda$ while $M/N$ remains constant, the cluster bias $\log w_i$ must remain unchanged:
\begin{equation*}
   \log w_i 
      = \left(\frac{\partial\log W(\lambda \mathbf{n})}{\partial (\lambda n_i)}\right)
      = \left(\frac{\partial\log W(\mathbf{n})}{\partial n_i}\right) .
\end{equation*}
This requires
\begin{equation*}
   \log W(\lambda \mathbf{n}) = \lambda \log W(\mathbf{n})   
\end{equation*}
and states that $\log W(\mathbf{n})$ is homogeneous function of $\mathbf{n}$ of degree 1.

\subsection{Derivation of Fundamental Eq.\ (\ref{PF_fundamental})}
Take the log of Eq.\ (\ref{mpd})
\begin{equation*}
   \log\frac{\nn_i}{N} = \log \ww_i - \beta i - \log q, 
\end{equation*}
multiply both sides by $\nn_i$ and sum over all $i$:
\begin{equation*}
   \sum \nn_i \log\frac{\nn_i}{N} = 
   \sum \nn_i \log\ww_i - 
   \beta M -
   (\log q) N , 
\end{equation*}
Using $S=\log\mathbf{\nn!}$, and the Euler relationship in  Eq.\ (\ref{logW}), we obtain
\begin{equation*}
   S + \log \tilde W = \beta M + (\log q) N  , 
\end{equation*}
which, combined with Eq.\ (\ref{Omega_S_W}), leads to the final result
\begin{equation}\label{app:PF:homo1}
   \log\Omega = \beta M + (\log q) N .  
\end{equation}
If we write this result as
\begin{equation}
   \log\Omega = M\left(\beta + \frac{N}{M} \log q\right)
              = N\left(\frac{M}{N}\beta + \log q\right),
\end{equation}
it will become clear that $\log\Omega$ is homogeneous of degree $1$ with respect to both $M$ and $N$ ($\beta$, $\log q$ and $M/N$ are all intensive), and satisfies  Euler's theorem:
\begin{equation}\label{app:PF:homo2}
   \log\Omega = M \parder{\log\Omega}{M}{}
                    + N \parder{\log\Omega}{N}{}. 
\end{equation}
Comparison with Eq.\ (\ref{app:PF:homo1}) leads to 
\begin{equation*} 
   \beta = \parder{\log\Omega}{M}{N},\quad
   \log q = \parder{\log\Omega}{N}{M} , 
\end{equation*}
which appears as Eq.\ (\ref{beta_q_fundamental}) in the text. 

\subsection{Derivation of Isothermal Condition Eq.\ (\ref{isoT})}
At equilibrium, a sol-phase system maximizes the microcanonical weight of the composite system. For a given partitioning of mass between the two phases, the sol phase maximizes its own microcanonical weight when it relaxes to the most probable distribution that corresponds to $M_\sol=M-M_\gel$, $N_\sol=N-1$. Therefore, we seek to maximize the partition function of the combined system,
\begin{equation*}
   \log\Omega_{M,N} = \log\Omega_{M_\sol,N-1} + \log\Omega_\gel .
\end{equation*}
Suppose that the giant cluster exchanges mass $\delta m$ with the sol. Such exchange must leave the overall partition function unchanged:
\begin{equation*}
   0 = \delta \log\Omega_{M_\sol,N-1} +  \delta  \log\Omega_\gel. 
\end{equation*}
We divide both sides by $\delta m$, and noting that $\delta M_\gel=-\delta M_\sol$, the result is
\begin{equation*}
   \frac{\delta \log\Omega_\sol}{\delta M_\sol}  =  \frac{\delta \log\Omega_\gel}{\delta M_\gel}.
\end{equation*}
Since the exchange is at constant number of clusters, we identify these derivatives as the temperatures in each phase and thus arrive at the isothermal condition:
\begin{equation*}
    \beta_\sol = \beta_\gel.
\end{equation*}
Since the number of clusters does not fluctuate during exchange reactions, pressures do not equalize and each phase may be at its own pressure.

\subsection{The Sol-Gel Boundary} 
Here we determine the cluster size that separates the sol from the gel. 
Suppose that $i_\sol$ is the \textit{maximum possible} cluster size in the sol region and $i_\gel$ is the \textit{minimum possible} cluster size in the gel region. Since $i_\gel$ and $i_\sol$ are neighboring masses, they meet the condition 
\begin{equation}\label{app:imax1}
      i_\gel - i_\sol = 1 . 
\end{equation}
If a mass $i_\gel$ is placed in the gel, we are left in the sol with $M_\sol=M-i_\gel$, $N_\sol=N-1$. Then the maximum possible cluster size in the sol is
\begin{equation}\label{app:imax2}
   i_\sol = M_\sol-N_\sol+1 =  M-N-i_\gel+2 .
\end{equation}
Combining with  Eq.\ (\ref{app:imax1}) we find
\begin{equation}
   i_\sol = \frac{M-N+1}{2},\quad
   i_\gel = \frac{M-N+3}{2}.
\end{equation}
Therefore, we identify the sol and the gel regions by the conditions:
\begin{align*}
      \sol:~~~~~~ & 1 \leq i \leq \imax/2;\\
      \gel:~~~~~~ & \imax/2 < i \leq \imax, 
\end{align*}
where $\imax=M-N+1$ is the maximum possible cluster in the $(M,N)$ ensemble. 

\subsection{Stockmayer's model} 
Here we derive the asymptotic form of Eq.\ (\ref{stockmayer}) for trifunctional monomers. Using the Stirling approximation, $x!\approx x^x \exp(-x)\sqrt{2\pi x}$, and $f=3$ the Stockmayer weight becomes
\begin{equation}\label{stockmayer:wi:thl}
      w_i = \frac{e^2}{\sqrt{\pi}} (12 i)^i (i+2)^{-i-5/2} . 
\end{equation}
Noting that 
\begin{equation}
   (i+2)^{-i-5/2} \sim i^{-i-5/2} e^{-2-5/i}
   \sim i^{-i-5/2} e^{-2},
\end{equation}
the previous result simplifies to
\begin{equation}\label{app:stock1}
   w_i =\frac{12^i}{\sqrt{\pi}}\,  i^{-5/2}  . 
\end{equation}
The corresponding selection bias $W(\mathbf{n})$ on arbitrary distribution $\mathbf{n}=(n_1,n_2,\cdots)$ is
\begin{equation}
   W(\mathbf{n}) = \prod_i w_i^{n_i} 
                 = \pi^{-N/2} 12^M \prod_i i^{-5n_i /2},
\end{equation}
where we have used $\sum n_i = N$ and $\sum in_i=M$. The constant factors raised to $N$ or $M$ are selection-neutral (they make identical contribution to all distributions of the ensemble) and may be dropped from the selection bias so that the Stockmayer bias reduces to 
\begin{equation}\label{stockmayer:scaling}
   w_i = i^{-5/2} .
\end{equation}
The expressions for the cluster bias in Eqs. (\ref{app:stock1}) and (\ref{stockmayer:scaling}) are functionally equivalent, i.e., they produce identical ensembles. 



\end{document}